\documentclass[aps,prl,twocolumn,groupedaddress]{revtex4}
\usepackage[pdftex]{graphicx}
\usepackage{epsfig}
\usepackage{amsmath,amssymb}

\begin{document}
\def\Nfour{\mathcal N\,{=}\,4}
\def\Ntwo{\mathcal N\,{=}\,2}
\def\Nc{N_{\rm c}}
\def\Nf{N_{\rm f}}
\def\x{\mathbf x}
\def\q{\mathbf q}
\def\v{\mathbf v}
\def\vs{v_{\rm s}}
\def\S{\mathcal S}
\def\half{{\textstyle \frac 12}}
\def\http       #1{\href{http://#1}{\tt http://#1}}
\def\arXiv      #1{\href{http://arxiv.org/abs/#1}{{\tt #1}}\,}
\def\Arxiv      #1 [#2]{\href{http://arxiv.org/abs/#1}{{\tt arXiv:#1 [#2]}}\,}

\title
{The wake of a quark moving through a strongly-coupled  $\mathcal{N}$=4
    supersymmetric Yang-Mills plasma}

\author{Paul~M.~Chesler}
\author{Laurence~G.~Yaffe}

\affiliation
    {Department of Physics, University of Washington, Seattle, WA 98195, USA}

\date{July 31, 2007}

\begin{abstract}
The energy density wake produced by a heavy quark moving through
a strongly coupled $\Nfour$ supersymmetric Yang-Mills plasma is
computed using gauge/string duality.
\end{abstract}

\pacs{}

\maketitle

{\it{Introduction}}.---The discovery that the quark-gluon plasma produced 
in heavy ion collisions at RHIC
behaves as a nearly ideal fluid \cite{Shuryak,Shuryak:2004cy}
has prompted much interest in understanding the dynamics of strongly coupled
non-Abelian plasmas.
Gauge/string duality \cite{Aharony:1999ti, Maldacena:1997re}
allows one to compute many observables
probing non-equilibrium dynamics of thermal $\Nfour$ supersymmetric
Yang-Mills (SYM) theory, including the rate of energy loss of a heavy
quark moving through an SYM plasma \cite{Herzog:2006gh}.
(See also Refs.~\cite {%
    Herzog:2006se,%
    Casalderrey-Solana:2006rq,%
    Friess:2006fk,%
    Gubser:2007nd,%
    Yarom:2007ni%
    }
and references therein.)
Energy transferred to the plasma from the moving quark will cause
the energy density of the plasma, in the vicinity of the quark,
to deviate from its equilibrium value.
That is, the moving quark will create an energy density ``wake''
which moves with it through the plasma.
The structure of this wake is of interest for studies of
jet quenching and jet correlations in heavy ion collisions
\cite{Shuryak:2006ii,Adler:2005ee}.
We evaluate this energy density perturbation,
$\langle \Delta T^{00}(x) \rangle$,
and display the resulting energy density wake in the case
of subsonic, transsonic, and supersonic motion~%
\footnote
    {
    See Refs.~\cite{Friess:2006fk,Gubser:2007nd,Yarom:2007ni}
    for related previous work.
    }.
    
{\it{Gravitational description.}}---According to gauge/string duality,
the addition of a massive quark to the $\Nfour$ SYM plasma is accomplished by
embedding a D7 brane in the AdS-Schwarzschild (AdS-BH) geometry and then
adding a string running from the D7 brane down to the black hole horizon
\cite{Karch:2002sh,Herzog:2006gh}.
The presence of the string perturbs the geometry via Einstein's equations.
The behavior of the metric perturbation near the AdS boundary encodes the
change in the SYM stress-energy tensor.
In the $\Nc \to \infty$ limit, the $5d$ gravitational constant
becomes parametrically small
and consequently the presence of the string acts as a small perturbation
on the AdS-BH geometry.
To obtain leading order results in $\Nf/\Nc$, we
write the full metric as $g_{\mu \nu} = g^{(0)}_{\mu \nu} + h_{\mu \nu}$,
where $g^{(0)}_{\mu \nu}$ is the metric of the AdS-BH geometry,
and then linearize the Einstein equations in the perturbation $h_{\mu \nu}$.
This gives
\begin{equation}
\label{lin1}
\Delta_{\mu \nu}^{ \alpha \beta} \, h_{\alpha \beta} = \kappa_5^2 \; t_{\mu \nu} \,,
\end{equation}
where $\Delta_{\mu \nu}^{ \alpha \beta}$ is a second order linear differential operator,
$\kappa_5^2 \equiv \frac{4 \pi^2 L^3}{\Nc^2}$ with $L$ the AdS curvature radius,
and $t_{\mu \nu}$ is the five dimensional stress-energy tensor
of the trailing string solution \cite{Herzog:2006gh}.
The boundary value of the metric perturbation acts as a source
for the SYM stress-energy tensor via the relation \cite{Skenderis:2000in}
\begin{equation}
\label{bdaction1}
    \langle {T^\mu}_\nu \rangle
    = 2\, \frac{\delta S_{\rm B}}{\delta {\overline h_\mu}\vphantom{h}^{\nu}}
    \bigg |_{\overline h_\mu\vphantom{h}^\nu= 0} \,,
\end{equation}
where $S_{\rm B}$ is the on-shell gravitational boundary action with
$\overline h_\mu\vphantom{h}^\nu$
the boundary value of the metric perturbation.

We choose a coordinate system such that the metric of the AdS-BH geometry is
\begin{equation}
\label{metric}
    ds^2 = \frac{L^2}{u^2}
    \left (-f(u) \, dt^2 + d \mathbf x^2 + \frac{du^2}{f(u)} \right ) ,
\end{equation}
where $f(u) \equiv 1-(u/u_h)^4$.
The event horizon is located at $u=u_h$,
with $T = (\pi u_h)^{-1}$ the temperature of the SYM plasma.
The boundary of the AdS-BH spacetime is at $u{=}0$.
The geometry is translationally
invariant in the four Minkowski space directions
($t,\mathbf x$),
so one may perform a spacetime Fourier transform
and work with mode amplitudes
$h_{\mu \nu}(u;\omega,\q)$.


{\it{Gauge invariants.}}---The $4d$ $\Nfour$ SYM stress-energy tensor 
is traceless and conserved,
and consequently contains five independent degrees of freedom.
This contrasts with the fifteen degrees of
freedom contained in the metric perturbation $h_{\mu \nu}$.
However not all of these degrees of freedom are physical.
The linearized field equations are invariant under coordinate
transformations $x^{\mu} \rightarrow x^{\mu} + \xi^{\mu}$,
where $\xi^{\mu}$ is an arbitrary infinitesimal vector field.
Under such transformations, the metric perturbation transforms as
\begin{equation}
\label{gaugetrans}
    h_{\mu \nu} \rightarrow
    h_{\mu \nu}-D_{\mu} \, \xi_{\nu} - D_{\nu} \, \xi_{\mu} \,,
    \end{equation}
where $D_{\mu}$ is the covariant derivative with respect to
the background metric $g^{(0)}_{\mu \nu}$.
Consequently,
five components of $h_{\mu \nu}$
may be eliminated via gauge fixing.
As in electromagnetism,
this does not completely fix the gauge.
If one chooses to set
$h_{5\nu} = 0$, where $x_5\equiv u$ denotes the AdS radial coordinate,
then the residual gauge freedom allows one to eliminate five more components
of $h_{\mu\nu}$ on any single $u = \mbox{const.}$ hypersurface.
This limits the number of independent physical degrees of freedom
carried by $h_{\mu \nu}$ to five, matching that of the SYM stress tensor.

Useful gauge invariants may
be constructed out of linear combinations of the Fourier mode amplitudes
$h_{\mu \nu}(u;\omega,\q)$
\cite{Kovtun:2005ev},
and classified according to their behavior under spatial rotations.
There is one helicity zero gauge invariant linear combination,
and a pair each of helicity one and two invariants
\footnote
    {
    There are also gauge invariants involving components of $h_{\mu\nu}$
    plus its radial derivatives.
    These will not be required.
    }.
Rotation plus gauge invariance implies that
these five invariants satisfy decoupled equations of motion.
For later convenience, let $H_{\mu\nu}\equiv (u^2/L^2)\, h_{\mu\nu}$.
A short exercise using Eq.~(\ref{gaugetrans}) shows that
\begin{eqnarray}
\label{Zdef}
    Z(u;\omega,\q) &\equiv&
	q^2  H_{00} + 2 \omega q^i H_{i0}
	+ (\omega^2/q^2) \, q^i q^j H_{ij}
\nonumber\\ &+&
	\half \left [2 {-} f(u){-} \omega^2/q^2 \right]
	(q^2 H_{ii} {-} q^i q^j H_{ij}) \ \ \ \
\end{eqnarray}
is the helicity zero gauge invariant linear combination of metric perturbations,
unique up to multiplication by an arbitrary function of $u$.
(Repeated Minkowski spatial indices $i,j =1,2,3$ are implicitly summed.)
It is straightforward (but tedious) to work out the equation of motion for
$Z$ from the linearized field equations.
Doing so, and inserting the explicit form of the trailing-string
gravitational stress tensor
(also computed in Ref.~\cite{Friess:2006fk})
yields the second order ODE
\begin{equation}
    Z'' + A(u) \, Z' + B(u) \, Z = S(u) \,,
\label{eqm}
\end{equation}
where
\begin{eqnarray}
    A(u)
    &\equiv&
    \frac{1}{u}
    \left[
	1
	+
	\frac{uf'}{f}+
	\frac{24 \left(q^2 f{-}\omega ^2\right)}
	    { q^2\left(u f'{-}6 f\right)+6 \omega^2}
    \right] ,
\\
    B(u) &\equiv&
    \frac 1f
    \left[
	-q^2 + \frac{\omega^2}{f}
	- \frac {32 \, q^2 u^6 u_h^{-8}}{q^2(u f' {-} 6f) + 6 \omega^2}
    \right] ,
\\
    S(u) &\equiv&
     \frac {\kappa_5^2 \sqrt\lambda} {6\pi L^3 } \,
     \frac{q^2 \left(v^2{+}2\right){-}3 \omega ^2} {q^2 \sqrt{1{-}v^2}} \,
\nonumber\\ &\times& {}
     \frac{ u [q^4 u^8+48 i q^2 \omega  u_h^2 u^5-9 (q^2{-}\omega ^2)^2 u_h^8]}
	  {f \left(f q^2+2 q^2-3 \omega ^2\right) u_h^8} \>
\nonumber\\ &\times& {}
	 2 \pi \delta(\omega {-} \v \cdot \mathbf q) \,
	 \, e^{- i \omega \> x_{\rm string}(u)} \,.
\label{eq:S}
\end{eqnarray}
Here $\lambda$ is the t' Hooft coupling,
$\v$ is the quark velocity,
and
\begin{equation}
    x_{\rm string}(u)
    \equiv
    \frac{u_h}{2}
    \left[
	\tan^{-1}\Bigl(\frac{u}{u_h}\Bigr)
	+ \half \log \Big( \frac{u_h{-} u}{u_h{+}u} \Big)
    \right]
\end{equation}
is the trailing string profile \cite{Herzog:2006gh}.

In a gauge in which $h_{5 \alpha} {=} 0$ for all $\alpha$,
the change in the
energy density is given by \cite{Skenderis:2000in}
\begin{equation}
\label{bndenergy}
    \langle \Delta T^{00} \rangle = \frac{2 L^3 }{\kappa_5^2} \,
    H_{00}^{(4)} \,,
\end{equation}
where
$H_{00}^{(4)}$ is the quartic term in the expansion of
$H_{\mu \nu} \equiv (u^2/L^2) \, h_{\mu \nu}$
at the boundary
\footnote
   {
   There is a subtle issue here which we are glossing over.
   The result (\ref{bndenergy}) was derived
   for the case of pure gravity and does not directly take into account 
   contributions to the boundary action from the D7 brane, which has
   support all the way to the boundary of the AdS-BH geometry.
   With a finite quark mass, neglecting the D7 brane is inconsistent 
   with Einstein's field equations, as the string stress tensor is not
   conserved at the end of the string (which is in the bulk).  For a large 
   quark mass $M$, the trailing string will deform the D7 brane 
   over length scales $\sim 1/M$.  However just as in HQET, in the limit
   $M \rightarrow \infty$, one may write down an effective gravitational theory 
   for the string/brane system.  To leading order the D7 brane (in the bulk)
   can be neglected and the perturbation to its $5d$ stress tensor can be approximated
   with that of the string near the boundary.  Doing so, one finds additional corrections 
   to Eq.~(\ref{bndenergy}) which have delta function support at the location of the quark.
   }.
A simple connection
between $Z$ and the energy density may be found by
considering the behavior of $Z$ and $H_{\mu \nu}$ near $u=0$.
Substituting a power series expansion into the linearized
equations of motion (\ref{lin1}),
one finds that near the boundary $Z$ and $H_{\mu \nu}$ have the forms
\begin{align}
    Z(u) &=   Z_{(3)} \, u^3 +   Z_{(4)} \, u^4+ \cdots,
    \\
    H_{\mu \nu}(u)&=  H_{\mu \nu}^{(3)} \, u^3 + H_{\mu \nu}^{(4)} \, u^4
    + \cdots .
\end{align}
Moreover,
the three combinations $q^i \, H_{i0}^{(4)}$,
$q^2 \, H_{ii}^{(4)}$, and
$q^i q^j H_{ij}^{(4)}$
are all determined by $H_{00}^{(4)}$.
Substituting the coefficients $H_{\mu \nu}^{(4)}$ into Eq.~(\ref{Zdef})
and solving for $H_{00}^{(4)}$ yields
\begin{eqnarray}
H_{00}^{(4)} = \frac{2 q^2}{3 (q^2 {-} \omega^2)^2} \left(Z_{(4)} {-} \mathcal A \right )
\end{eqnarray}
where
\begin{align}
    \mathcal A &=
    \frac {\kappa_5^2 \sqrt\lambda} {8\pi L^3 } \>
    \frac{i \omega
	\left[
	    q^2 \left(5 v^2{+}1\right) -3 \left(v^2{+}1\right) \omega^2
	\right]}
	{q^2 u_h^2 \sqrt{1{-}v^2}}
    \nonumber \\
     &\times (2 \pi) \, \delta(\omega {-} \mathbf v \cdot \mathbf q).
\end{align}
The temperature dependent perturbation in the energy density
(which we focus on below) is therefore given by
\begin{align}
\label{FTenergy}
    \mathcal E(\omega,\mathbf q) &\equiv
    \langle \Delta T^{00} \rangle -
    \langle \Delta T^{00} \rangle_{T{=}0}
\nonumber\\
    &=
    \frac{4 q^2 L^3}{3 \kappa_5^2 (q^2{ -} \omega^2)^2}
    \left( \Delta Z_{(4)} {-} \mathcal A \right ) \,,
\end{align}
where $\Delta Z_{(4)}$ is the difference between $Z_{(4)}$ evaluated
at temperature $T$ and zero temperature.


{\it{Asymptotics and numerics.}}---We 
solve Eq.~(\ref{eqm}) with a Green's function $G(u,u')$ constructed out
of homogeneous solutions,
\begin{equation}
    G(u,u') = g_{<}(u_{<}) \> g_{>}(u_>) \big/ W(u') \,,
\end{equation}
where $W(u)$ is the Wronskian of $g_<$ and $g_>$.
The appropriate homogeneous solutions are dictated by the boundary conditions.
The differential operator in
(\ref{eqm}) has singular points at $u=0$ and $u=u_h$ with exponents 0 and 4,
and $\pm i \omega u_h/4$, respectively.
Vanishing of
the metric perturbation near the boundary requires that
$g_{<}(u) \sim u^4$ as $u\to0$,
while the requirement that the black hole not radiate \cite{Son:2002sd} implies that
$g_{>}(u) \sim (u{-}u_h)^{-i \omega u_h/4}$ near the horizon.
The overall
normalization of $g_<$ may be fixed by requiring $\lim_{u\to0} g_{<}(u)/u^4 \equiv 1$.
Zero temperature solutions to Eq.~(\ref{eqm}) are easily found
and involve modified Bessel functions.
A short exercise leads to
\begin{align}
\label{grnsfcnrep}
    \Delta Z_{(4)}& =
    \int_{0}^{u_h} du \> \Bigr \{\frac{g_{>}(u)}{W(u)} \> S(u)
\nonumber \\
    &+
    J(u)\,\frac{q^2{-}\omega^2}{8 u} \,
    K_2 \Big (\frac{u\sqrt{q^2 {-} \omega^2}}{f(u)} \Big)
    S_{0}(u) \Bigr \} \,,
\end{align}
where $J \equiv (f - u f')/f^2$ is a Jacobian factor
and $S_0(u)$ is the source $S(u)$ evaluated at $T=0$.

For very large or small momneta (compared to $T$),
one may find explicit asymptotic expressions for the homogeneous solutions
and derive the asymptotic
behavior of $\mathcal E (\omega,\mathbf q)$.
The large momentum limit is
physically identical to the low temperature limit,
so expanding Eq.~(\ref{eqm}) about $u_h = \infty$
allows one to extract the large $q$ asymptotics.
For a fixed ratio $r \equiv \omega/q$,
the linear differential operator on the left side of Eq.~(\ref{eqm})
differs from its $T{=}0$ limit only by terms of relative order $T^4/q^4$,
while the source $S$ equals
$
    S_0 \left (1+i \omega u^3/3 u_h^2 \right )
$
up to $O(T^4/q^4)$ corrections.
Defining for convenience
\begin{equation}
    s(\omega, \mathbf q) \equiv \frac{ \sqrt{\lambda } } {2 \pi \sqrt{1{-}v^2}}
    \> (2 \pi) \delta(\omega {-} \mathbf v \cdot \mathbf q) \,,
\end{equation}
we find the large momentum asymptotic behavior
\footnote
    {
    Our asymptotic forms (\ref{near}) \& (\ref{far})
    agree with Refs.~\cite{Friess:2006fk,Yarom:2007ni,Gubser:2007nd}.%
    \hspace*{-20pt}%
    }%
\begin{align}
\label{near}
    \mathcal E(\omega ,\mathbf q) = s(\omega,\mathbf q) \,
    \frac{i \omega
	[ \left(5{-}11 v^2\right) q^2{+}3 \left(3 v^2{-}1\right) \omega^2 ]
	} {9 u_h^2\left(q^2{-}\omega^2\right)^2} \,,
\end{align}
up to relative corrections suppressed by $1/q^2$.
For small momentum, we expand Eq.~(\ref{eqm}) in powers
of $q$ with the ratio $\omega/q$ fixed.
The linear differential operator on the left side of Eq.~(\ref{eqm})
has a smooth $q\to0$ limit with $O(q^2)$ corrections,
but relative $O(q)$ terms appear in the source $S$ and the Green's function
(via the incoming boundary condition at the horizon).
As $q \rightarrow 0$, we find
\begin{align}
\label{far}
    \mathcal E (\omega,\mathbf q)
    =
    \frac{3 s(\omega,\q)}{\left(1{-}3 r^2\right)}
    \biggl [
	\frac{   r \left(1{+}v^2 \right)}{ i q \, u_h^2 }
	+\frac{ r^2 \left(2{+}v^2{-}3 r^2\right)}{u_h \left(1{-}3 r^2\right) }
    \biggr ],
\end{align}
up to relative corrections suppressed by $q^2$.

To compute the energy density $ \mathcal E (x)$
for a given quark speed $v\, {=} \,|\v|$,
we use the cylindrical symmetry of the source (\ref{eq:S}) to reduce
the spacetime Fourier transform to a
two dimensional integral over $q_\| \equiv \hat \v \cdot \q$ and
$q_\perp \equiv |\hat \v \times \q|$, which is performed numerically.
For each value of $q_\|$ and $q_\perp$, the Fourier amplitude
$\mathcal E (v q_\|, q_\|, q_\perp)$,
as given by Eq.~(\ref{FTenergy}),
is evaluated by
numerically integrating the homogeneous differential equation (\ref{eqm})
(without source)
outward from the horizon to find $g_>(u)$,
and then evaluating numerically the radial integral (\ref{grnsfcnrep})
to find $\Delta Z_{(4)}(v q_\|, q_\|, q_\perp)$.
(The $u$ dependence of the Wronskian can be computed analytically.)


\begin{figure}
\includegraphics[scale=0.21]{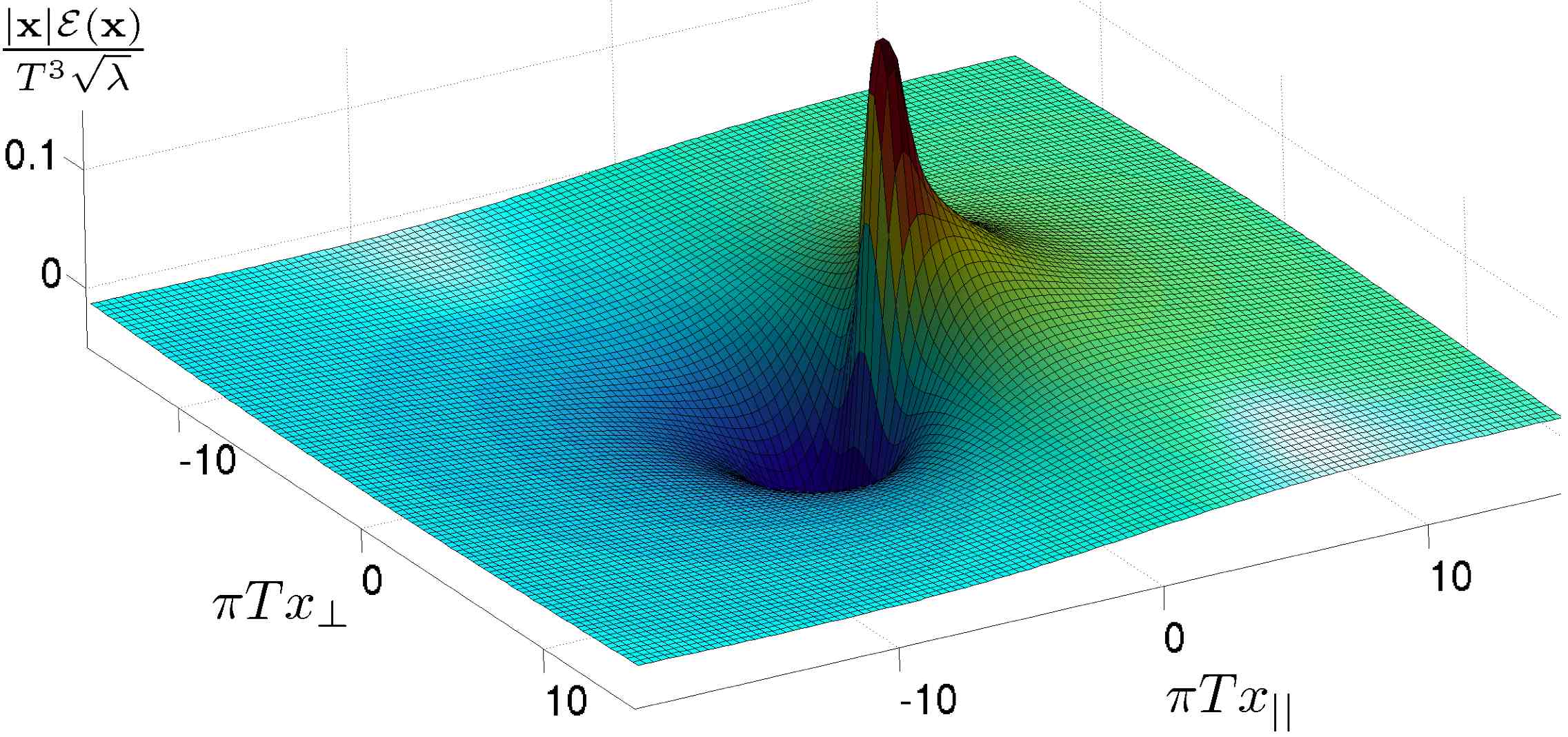}\!\!\!\!
\vspace*{-4pt}
\caption{\label{v025}
Plot of $|\x| \, \mathcal E(\x)/(T^3 \sqrt{\lambda})$ for $v = 1/4$,
with the zero temperature and near zone (\ref{near}) contributions removed.
Note the absence of structure in the region $|\x| \gg 1 / \pi T$.}
\end{figure}

{\it{Results and discussion.}}---For small distances
$d \equiv |\x{-}\v t| \ll 1/T$
away from the moving quark,
the dominant contributions to the energy density
come from momenta $q \gg T$.
The leading short distance behavior is temperature independent;
$T{=}0$ conformal invariance implies that the energy density
scales like $1/d^4$.  The first temperature
dependent near zone contribution to the energy density comes from the
term (\ref{near}).  This yields a position space energy density
which scales like $T^2/d^2$ in the vicinity of the quark.

\begin{figure}
\includegraphics[scale=0.21]{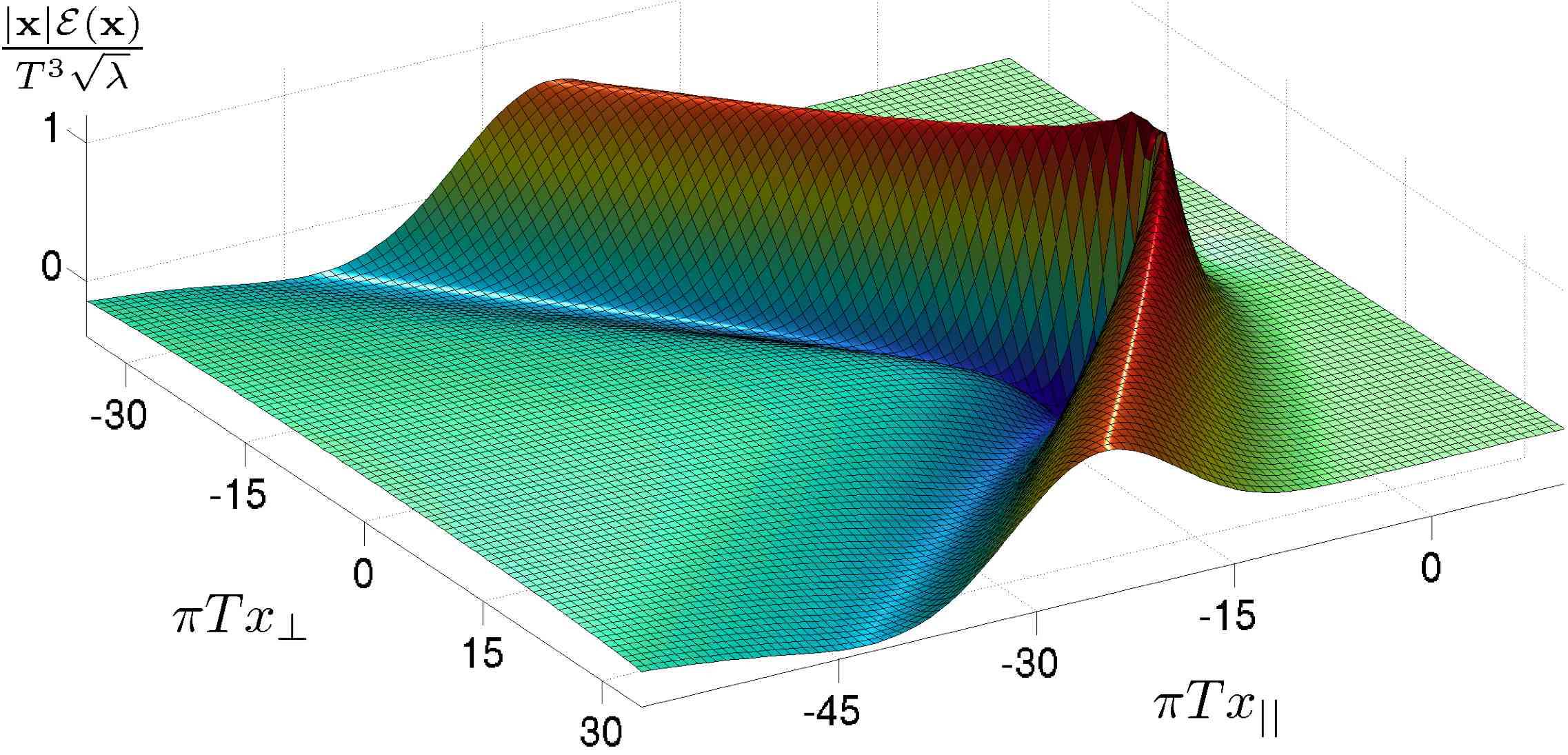}\!
\vspace*{-4pt}
\caption{\label{v075}
Plot of $|\x| \, \mathcal E(\x)/(T^3 \sqrt{\lambda})$ for $v = 3/4$,
with the $T{=}0$ and near zone (\ref{near}) contributions removed.
A Mach cone is clearly visible, with an opening half-angle
$
    \theta \approx 50^\circ
$.}
\end{figure}

\begin{figure}
\includegraphics[scale=0.21]{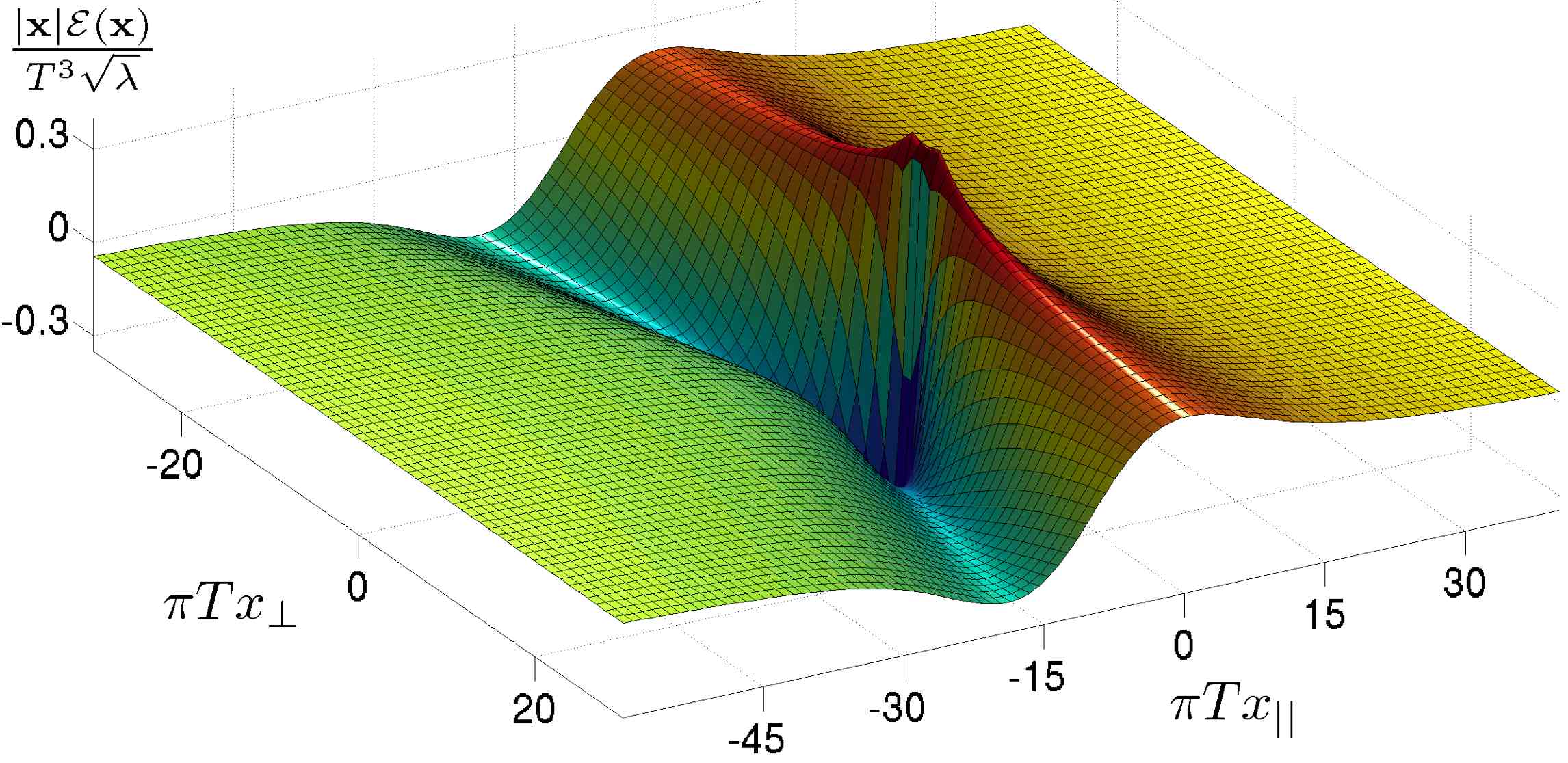}\!\!\!\!
\vspace*{-4pt}
\caption{\label{vs}
Plot of $|\x| \, \mathcal E(\x)/(T^3 \sqrt{\lambda})$ for $v = 1/\sqrt{3}$,
with the $T{=}0$ and near zone (\ref{near}) contributions removed.
A planar shock front perpendicular to the quark velocity is evident.
}
\end{figure}

Figures~\ref{v025}--\ref{vs} show real space plots of $\mathcal E(x)$
for quark velocities $v=1/4$, $3/4$, and $1/\sqrt{3}$, respectively,
with the near zone contribution (\ref{near}) removed
(in order to highlight the intermediate and far zone structure)
\footnote
    {
    In making these plots, spatial grids with resolutions
    $\Delta x \approx (0.3$--$1)/(\pi T)$ were used.
    This necessarily limits the fidelity of these plots
    at distances $|\x| \lesssim \Delta x$ from the quark.
    }.
In these plots the quark mass is infinite and
the quark's location, at the time shown, is $\x =0$.
Since $\Nfour$ SYM is a conformal theory,
the speed of sound is $1/\sqrt 3$.
Hence Fig.~1 shows subsonic motion,
Fig.~2 shows supersonic motion, and Fig.~3 is precisely at the speed of sound.
For all three velocities we observe a net surplus of energy in front of the
quark and a net deficit behind the quark.
This may naturally be interpreted as plasma being pushed and displaced
by the quark, just like the water displacement produced by a moving boat.

The most striking feature in these plots is the formation of a
conical energy wake, or sonic boom,
for velocities $v \ge \vs=1/\sqrt{3}$.
A textbook constructive interference argument shows that
a projectile moving supersonically,
in any fluid,
should produce a Mach cone
with an opening half-angle given by
$\sin \theta = \vs/v$ (where $\tan \theta \equiv -x_\perp/x_\|$).
For $v=3/4$, this is $50.3^\circ$.
For the transsonic $v{=}\vs$ case shown in Fig.~\ref{vs},
we see an energy wake along the plane front $x_\|=0$,
while for $v=3/4$ the wake is concentrated, as expected,
along a $50^\circ$ cone.
From Fig.~\ref{v075}, one may see that the shock wave has
a width $\approx 10/\pi T$ and broadens with increasing distance.
This behavior is to be expected in a viscous fluid where
sound waves are damped and diffusive.  Furthermore,
the amplitude of the 
shock front decreases slightly faster than $1/d$. 
If the shock front did not broaden with increasing distance,
then conservation of energy would imply a $1/d$ decrease of
the intensity of the shock wave in the far zone.

It is instructive to compare the long wavelength limit (\ref{far}) of the energy
density with the behavior predicted by
linearized hydrodynamics, in which the energy
density perturbation $\mathcal E$ satisfies the diffusive wave equation
\begin{equation}
\label{linhydro}
    \left (
	- \partial_t^2
	+\gamma \nabla^2\partial_t
	+\vs^2 \nabla^2
    \right ) \mathcal E = \rho \,.
\end{equation}
Here $\rho$ is an effective source which depends on
dynamics in the near zone, and
$\gamma \equiv \frac{4 \eta} {3(\epsilon{+}p)}$
is the sound attenuation constant.
(See also Ref.~\cite{Casalderrey-Solana:2006sq}.)
In strongly coupled $\Nfour$ SYM,
$\gamma = (3\pi T)^{-1}$ \cite{Policastro:2002tn}.
Fourier transforming
and expanding in powers of momentum (at fixed $r=\omega/q$) with
$\rho =q \rho_1 + q^2\rho_2  + \cdots$,
one has
\begin{equation}
\label{hydro}
    \mathcal E
    =
    -\frac{3 \rho_1} {(1{-}3 r^2)\, q}
    - \frac{9 i  r \, \gamma\, \rho_1 + 3 (1{-}3 r^2) \rho_2}{( 1{-}3 r^2)^2}
    + \cdots \,.
\end{equation}
Comparing the above with the sound poles in Eq.~(\ref{far}),
we see that our result for the far zone energy density agrees with
linear hydrodynamics provided $\gamma$ has the expected
value of $(3\pi T)^{-1}$ and
we identify%
\begin{equation}
\label{adssource}
    \rho(\mathbf x,t) = -\frac{\sqrt{\lambda} \, u_h^{-2}}{2 \pi \sqrt{1{-}v^2}}
     \> (1{+}v^2) \, \partial_t \, \delta^{3}(\mathbf x {-} \mathbf v t),
\end{equation}
up to higher derivative corrections.
This may be compared with the source obtained from the energy momentum
conservation equation
$\partial_\mu T^{\mu \nu} = F^\nu$,
where $F^\mu  = f^{\mu} \, \delta^3( \mathbf x {-} \mathbf v t) \sqrt{1{-}v^2}$
and $f^{\mu}$ is the external force (or minus the drag force)
acting on the quark.
Using the equations of
linearized hydrodynamics, one may calculate $\rho$ in terms of $F^{\nu}$.
To leading order (in derivatives) one finds
\begin{equation}
\label{classicalsource}
    \rho = \nabla \cdot \mathbf F - \partial_t \, F^0.
\end{equation}
Inserting the drag force  computed in Ref.~\cite{Herzog:2006gh}
precisely reproduces the result (\ref{adssource}). 
Unsurprisingly, higher derivative terms in the source are not
reproduced by linear hydrodynamics.

Natural extensions of this work include the calculation of energy flux
and the inclusion of a non-zero chemical potential.
As this work was nearing completion,
we learned of a similar study \cite{Gubser:2007} of the energy density
of a moving quark.
We thank the authors of Ref.~\cite{Gubser:2007},
as well as A. Karch, C.P. Herzog, and D.T. Son, for useful discussions.
This work was supported in part by the U.S. Department
of Energy under Grant No.~DE-FG02-96ER40956.
%
\bibliography{refs}%
\end{document}